\documentclass[aps,twocolumn,pra,superscriptaddress,showpacs,tightenlines]{revtex4}
\usepackage{amssymb}
\usepackage{amsmath}
\usepackage{graphicx}
\usepackage{epsfig}
\usepackage{txfonts}
\usepackage{subfigure}
\usepackage{amsfonts}
\usepackage{CJK}
\usepackage{mathrsfs}
\begin{document}

\begin{CJK*}{GBK}{song}
\title{Complementarity  between micro-micro and micro-macro entanglement in a Bose-Einstein condensate with two Rydberg impurities }
\author{Zhen Li}
\author{Yu Han}
\author{Le-Man Kuang\footnote{Author to whom any correspondence should be
addressed. }\footnote{ Email: lmkuang@hunnu.edu.cn}}
\affiliation{Key Laboratory of Low-Dimensional Quantum Structures
and Quantum Control of Ministry of Education,  Department of Physics and  Synergetic Innovation Center for Quantum Effects and Applications,
Hunan Normal University, Changsha 410081, China}
\date{\today}

\begin{abstract}
We theoretically study complementarity  between micro-micro and micro-macro entanglement in a Bose-Einstein condensate with two Rydberg impurities. We investigate quantum dynamics of micro-micro and micro-macro entanglement in the micro-macro system.  It is found that strong micro-macro  entanglement  between Rydberg impurities and the BEC can be generated by the use of  initial micro-micro entanglement between two Rydberg impurities, which acts as the seed entanglement to create micro-macro entanglement.  We demonstrate a curious complementarity relation between   micro-micro  and    micro-macro entanglement, and find   that the complementarity property can be sustained to some extend even though in the presence of the BEC decoherence.
\end{abstract}
%03.67.-a: Quantum information
%03.65.Ta: Foundations of quantum mechanics
%03.65.Yz: Decoherence; open systems; quantum statistical methods

\maketitle \narrowtext
\end{CJK*}

\section{\label{Sec:1}Introduction}

Quantum entanglement is a distinctive feature of quantum mechanics that lies at the core of many new quantum applications in the emerging science
of quantum information. In particular, it is an open challenge for fundamental physics that to reveal the quantum entanglement of a microscopic-macroscopic system \cite{1,JQ2009,2,3,4,5,6,7,JQ2015,JQ2016,JQL2016,JQ2012,8,9}. This endeavor could
contribute to challenge the observability of quantum features at the macroscopic level, which is one of the very fascinating open problems in quantum physics.
The difficulties inherent in such a question are manifolds, and they are related  not only to quantum decoherence induced by the surrounding environment \cite{10,FD2008,NS2011,YJ2019,11,QS2017,JB2017,JBY2017,12,LM2007,13}, but also to a measurement precision sufficient to observe quantum effects at such macroscales.

On the other hand, complementarity is one of the most characteristic properties of quantum mechanics \cite{14,15,16}, which distinguishes the quantum world from the classical one.
Quantum entanglement in composite quantum systems is a uniquely quantum resource with no classical counterpart. An interesting problem to ask is whether quantum entanglement can be incorporated into the principle of complementarity. Some authors have explored this question and obtained some interesting results, such as the complementarities between distinguishability and entanglement \cite{17}, between coherence and entanglement \cite{18}, and between local and nonlocal information \cite{19}, etc. Additionally, some complementarity relations in multi-qubit pure systems are also observed such as the relationships between multipartite entanglement and mixedness \cite{20,21}, and between the single-particle properties and the $n$ bipartite entanglements \cite{22}. In this paper, we add the complementarity between micro-micro and micro-macro entanglement into the complementarity list on quantum mechanics.

A Bose-Einstein condensate (BEC) with Rydberg impurities \cite{23,24,25,26,27}  presents a totally new platform to study  micro-micro and micro-macro entanglement where microscopic impurities meets a macroscopic matter, the BEC. As
the interaction among Rydberg atoms can be tailored by electric fields and microwave fields  \cite{28,29} while the BEC allows for an extremely precise control of interatomic interactions by manipulating $s$-wave scattering length \cite{30,31}, they provide a new tool for many-body quantum physics of hybrid quantum systems \cite{32},  consisting of microscopic-macroscopic systems.

In this paper, we study complementarity  between micro-micro and micro-macro entanglement in a Bose-Einstein condensate with two Rydberg impurities. We investigate quantum dynamics of  micro-micro and micro-macro entanglement in  the impurities-doped BEC  system.  It is found that strong micro-macro  entanglement  between Rydberg impurities and the BEC can be generated by the use of  initial micro-micro entanglement between two Rydberg impurities, which acts as the seed entanglement to create micro-macro entanglement. It is shown that the micro-macro system under our consideration exhibits not only micro-micro entanglement collapse and revival (ECR) between two Rydberg impurities, but also micro-macro ECR between  Rydberg impurities and the BEC.  We demonstrate a curious complementarity relation between the micro-micro and micro-macro entanglement, and find  that micro-micro entanglement can perfectly transfer into the micro-macro entanglement.

The remainder of this paper is organized as follows. In Sec.\uppercase\expandafter{\romannumeral2}, we introduce the impurities-doped BEC model consisting of the BEC and two Rydberg impurities. We obtain an analytical solution of the impurities-doped BEC model. In Sec.\uppercase\expandafter{\romannumeral3}, we investigate quantum dynamics of micro-micro entanglement between two Rydberg impurities, and the influence of the BEC decoherece. It is shown that there exists the ECR of  micro-micro entanglement between two Rydberg impurities, the BEC decoherence suppresses the revival of  micro-micro entanglement.  In Sec.\uppercase\expandafter{\romannumeral4}, we demonstrate the complementarity between micro-micro and micro-macro entanglement by calculating  micro-macro entanglement, and obtain an exact complementarity between micro-micro and micro-macro entanglement. It is indicated   that the complementarity property can be sustained to some extend even though in the presence of the BEC decoherence. Finally, Sec.\uppercase\expandafter{\romannumeral5} is devoted to some concluding remarks.

\section{\label{Sec:2}Model Hamiltonian}

The microscopic-macroscopic system under our consideration consists  of a BEC and  two localized Rydberg impurity atoms  immersed in the BEC. The two separated Rydberg impurities  are frozen in place and they interact with each other via a repulsive van der waals interaction \cite{28,29}. The relevant internal level structure for each Rydberg atom is given by the atomic ground state $|0\rangle$ and the excited Rydberg state $|1\rangle$, which form a two-level system. The Hamiltonian of two Rydberg impurities in the absence of the external laser field  \cite{29} is given by
\begin{eqnarray}
H_R=\frac{\omega}{2}(\sigma_z^1+\sigma_z^2)+J\sigma_z^1\sigma_z^2,
\end{eqnarray}
where $\omega$ is the transition frequency between two internal states of each Rydberg impurity atom, the second term accounts for the van der Waals  interaction between the Rydberg impurities with the coupling strength $J=C_6/R^6$ where $R$ is the distance between two  localized Rydberg impurities, $C_6\propto \bar{n}^{11}$ with $\bar{n}$ being the principal quantum number of the Rydberg excitation.  We have set $\hbar=1$ in Hamiltonian (1) and through out the paper.

The Hamiltonian of a BEC confined in a trapping potential  is given by
\begin{eqnarray}
H_B=\int d\textbf{x} \Psi^{\dag}(\textbf{x}) \left[-\frac{1}{2m}\nabla^2+V(\textbf{x}) + \frac{U}{2}\Psi^{\dag}(\textbf{x})\Psi(\textbf{x})\right]\Psi(\textbf{x}),
\end{eqnarray}
where $\Psi(\textbf{x})$ is the annihilator field operator of the BEC  at the position $\textbf{x}$, $V(\textbf{x})$
is the external trapping potential, $U$ is the inter-atomic interaction strength, and $m$ is the mass of an atom.
Assume that  the BEC is trapped in a deep potential,  we can use the single-mode approximation $\Psi(\textbf{x})\approx a \phi(\textbf{x})$ to describe the BEC with $a$ and $\phi(\textbf{x})$ being the annihilation operator and the mode function of the condensate, respectively. Then, the Hamiltonian of the BEC (2) can be written as the following Kerr-interaction form
\begin{eqnarray}
H_B=\omega_b a^\dag a+\chi a^{\dag}a^{\dag} aa,
\end{eqnarray}
where the mode frequency $\omega_b$ and the coupling constant $\chi$ are defined by
\begin{eqnarray}
\omega_b&=&\int d\textbf{x} \left[-\frac{1}{2m}|\nabla \phi(\textbf{x})|^2+V(\textbf{x})|\phi(\textbf{x})|^2\right], \\
\chi&=&\frac{U}{2}\int d\textbf{x} |\phi(\textbf{x})|^4.
\end{eqnarray}

%%%%%%%%%%%%%%%%%%%%%%%%%%%%%%%%%%%%%%%%%%%%%%%%%%%%%%%%%%%
\begin{figure}[htp]
\includegraphics[width=8.5cm,height=3.0cm]{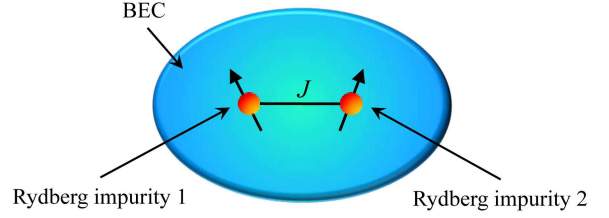}
\caption{Schematic diagram of two Rydberg impurities immersed in a Beose-Einstein condensate. Here $J$ denotes the van der Waals  interaction between  two Rydberg impurities.}
\end{figure}
%%%%%%%%%%%%%%%%%%%%%%%%%%%%%%%%%%%%%%%%%%%%%%%%%%%%%%%%%%%

The two Rydberg impurities interact with the BEC via coherent collisions. The impurity-BEC interaction Hamiltonian can be described as
\begin{eqnarray}
H_I=\frac{\lambda}{2}(\sigma_z^1+\sigma_z^2)a^\dag a,
\end{eqnarray}
and $\lambda$ is the interaction strength.

Hence, The Hamiltonian of the total system including the two  Rydberg impurities  and the BEC is given by
\begin{align}
H=&H_R + H_B +H_I,
\end{align}
which is a diagonal Hamiltonian with the following eigenvalues and eigenstates
\begin{eqnarray}
E_{ijn}&=&\frac{1}{2}\omega\left[(-1)^{i} + (-1)^{j}\right] + (-1)^{i+j}J \nonumber \\
&&+\frac{1}{2}\lambda\left[(-1)^{i} + (-1)^{j}\right]n +\omega_bn + \chi n(n-1), \\
|\psi\rangle_{ijn}&=&|i j n\rangle,
\end{eqnarray}
where  $|i j n\rangle=|i\rangle\otimes|j\rangle\otimes|n\rangle$ with $|i\rangle (|j\rangle)$ is a eigenstate of $\sigma_z^i (\sigma_z^j)$ with $i=0,1 (j=0,1)$, and $|n\rangle$ is a Fock state with $n=0, 1, 2, \cdots, \infty$.

In what follows we will use the total Hamiltonian (7) to study  quantum dynamics of micro-micro and micro-macro entanglement in the impurity-doped BEC system. We will demonstrate  the complementarity  between micro-micro and micro-macro entanglement, and uncover the physical mechanism behind the micro-micro and micro-macro complementarity in the dynamical evolution.

\section{\label{Sec:3} Micro-micro entanglement between two Rydberg impurities}

In this section, we want to explore dynamical characteristics of micro-micro entanglement between two Rydberg impurities and show the existence of  entanglement  collapse and revival (ECR) phenomenon for the two Rydberg impurities in the dynamical evolution. Concretely, we investigate  micro-micro entanglement between the two Rydberg impurities when  the two Rydberg impurities are initially in a Bell-type state while the BEC is initially in a coherent state  and an arbitrary pure state, respectively.

\subsection{The coherent state case}
We assume that the two Rydberg impurities  initially is a Bell-type state $(\cos\theta|00\rangle+\sin\theta|11\rangle)$  with  $|0\rangle$ and $|1\rangle$ denoting the ground state and excited Rydberg state of each Rydberg impurity, respectively, and the initial state of the BEC is a coherent state $|\alpha\rangle=\exp(-|\alpha|^2/2)\sum_{n=0}^{\infty}\alpha^n/\sqrt{n!}|n\rangle$. Then, the initial state of the micro-macro system under our consideration can be written as
\begin{eqnarray}
|\psi(0)\rangle=(\cos\theta|00\rangle+\sin\theta|11\rangle)\otimes|\alpha\rangle,
\end{eqnarray}

Making use of the Hamiltonian of the total system (7), we can get the state of the system at time $t\geq 0$
\begin{eqnarray}
|\psi(t)\rangle&=&[\cos\theta|00\rangle\otimes |\varphi_0(t)\rangle +\sin\theta|11\rangle\otimes|\varphi_1(t)\rangle],
\end{eqnarray}
where $|\varphi_0(t)\rangle$ and $|\varphi_1(t)\rangle$ are two generalized coherent states \cite{21,22,23,24}
\begin{eqnarray}
|\varphi_0(t)\rangle &=& e^{-\frac{|\alpha|^2}{2}} \sum_{n=0}^{\infty} e^{it\theta_0(n)}\frac{\alpha^n}{\sqrt{n!}}|n\rangle, \\
|\varphi_1(t)\rangle &=& e^{-\frac{|\alpha|^2}{2}} \sum_{n=0}^{\infty} e^{it\theta_1(n)}\frac{\alpha^n}{\sqrt{n!}}|n\rangle,
\end{eqnarray}
where we have introduced two running frequencies
\begin{eqnarray}
\theta_0(n)&=&\lambda n-\chi n(n-1)+\omega -\omega_{b}n-J, \\
\theta_1(n)&=&-\lambda n -\chi n(n-1) -\omega -\omega_{b}n-J.
\end{eqnarray}

From Eq. (9) it is easy to get the reduced density operator of the two Rydberg impurities
\begin{align}
\rho_R(t)=&\cos^{2}\theta|00\rangle\langle00|+\sin^{2}\theta|11\rangle\langle11|\nonumber\\
&+\xi(t)|00\rangle\langle11| +\xi^{*}(t)|11\rangle\langle00|,
\end{align}
where $ \xi(t)$ is defined by
\begin{equation}
 \xi(t)=\frac{1}{2}\sin{2\theta}\exp\left[2i\omega t-|\alpha|^2(1-e^{2i\lambda t})\right]
\end{equation}

We can use quantum concurrence  \cite{33}  to measure the amount of entanglement for an arbitrary quantum state of the two Rydberg impurities. The concurrence of an arbitrary quantum state of two qubits with a density operator $\rho_R(t)$ \cite{33}  is given by
\begin{eqnarray}
\mathcal{C}_1=\rm{max}\{0,\lambda_1-\lambda_2-\lambda_3-\lambda_4\},
\end{eqnarray}
where the $\lambda_i$ ($i=1,2,3,4$) are the square roots of the eigenvalues in descending order of the operator $R=\rho_R(t)(\sigma_y^1\otimes\sigma_y^2)\rho_R^{\ast}(t)(\sigma_y^1\otimes\sigma_y^2)$ with $\sigma_y$ being the Pauli operator in the computational basis. It ranges from $\mathcal{C}_1=0$ for a separable state to $\mathcal{C}_1=1$ for a maximally entangled state.

According to \cite{34}, if the time-dependent density matrix of a two qubit system can be expressed as
\begin{equation}
\rho(t)=\left(
  \begin{array}{cccc}
    w(t) & 0 & 0 & z(t)\\
    0 & x(t) & 0 & 0\\
    0 & 0 & x(t) & 0\\
    z^{\ast}(t) & 0 & 0 & y(t)\\
  \end{array}
\right)  ,
\end{equation}
one finds that the concurrence corresponding to this state is given by
\begin{equation}
\mathcal{C}_1(t)={\rm max}\{0,2|z(t)|-2x(t)\}.
\end{equation}

For the quantum state of two Rydberg impurities $\rho_R(t)$ given by Eq. (16), we have $x(t)=0$ and $z(t)=\xi(t)$. Hence we can obtain the quantum concurrence between two Rydberg impurities  with the following expression
\begin{equation}
\mathcal{C}_{1}(t)=\mathcal{C}_{1}(0)\exp\left\{|\alpha|^{2}\left[{\rm cos}(2\lambda t)-1\right]\right\},
\end{equation}
where $\mathcal{C}_{1}(0)$ is the initial quantum concurrence between two  Rydberg impurities  with the following expression
\begin{equation}
\mathcal{C}_{1}(0)=\big|\sin{(2\theta)}\big|.
\end{equation}

%\begin{figure}[htp]
%\includegraphics[width=8.0cm,height=6.0cm]{fig2.eps}
%\caption{Two Rydberg atoms concurrence versus time $t$ with different BEC initial coherent state parameter $\alpha$ in the situation $\omega_z=\omega_b=J=\chi=0.5$, $\theta=\pi/4$ and $\lambda=1$.}\label{fig2}
%\end{figure}

\begin{figure}[htp]
\includegraphics[width=8.0cm,height=6.0cm]{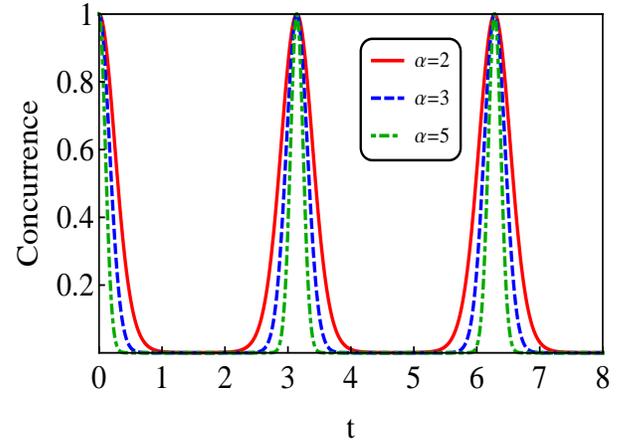}
\caption{Micro-Micro entanglement between two Rydberg impurities  denoted by the concurrence versus the scaled time $\tau$  for different values of the  initial-state parameter $\alpha$  with $\alpha=2,3$, and $5$, when the  two Rydberg impurities is initially in the Bell state with $\theta=\pi/4$. }\label{fig4}
\end{figure}

From Eq. (21) we can see that the entanglement dynamics between two Rydberg impurities depends on the initial entanglement $\mathcal{C}_{1}(0)$, the initial-state parameter  of the BEC, and the impurity-BEC interaction strength $\lambda$. It is independent of interaction between two Rydberg impurities $J$ and inter-atomic interaction in the BEC $\chi$. Obviously, without the initial entanglement between two Rydberg impurities, i.e., $\mathcal{C}_{1}(0)=0$, there does not exist entanglement dynamics between two  Rydberg impurities. In fact,  the initial entanglement   between two Rydberg impurities  $\mathcal{C}_{1}(0)$ is the maximal entanglement amount  of two Rydberg impurities in the process of dynamic evolution. On the other hand, Eq.(21) indicates that the entanglement dynamics between two Rydberg impurities is periodic with the evolution period $T=\pi/\lambda$, which is determined only by the impurity-BEC coupling strength $\lambda$. The stronger the impurity-BEC coupling strength is, the faster the entanglement oscillation in the dynamic evolution.

Eq. (21) indicates that one can control the micro-micro entanglement dynamics between two Rydberg impurities by changing  the amount of the initial micro-micro entanglement, the initial-state parameter  of the BEC, and the impurity-BEC interaction strength. In Fig. 2, we plot the evolution of the concurrence between two Rydberg impurities  with respect to the scaled time $\tau=\lambda t$ for different values of the initial-state parameter of the BEC $\alpha=2,3$ and $5$, when the two Rydberg impurities are initially in a Bell state with $\mathcal{C}_{1}(0)=1$, i.e., $\theta=\pi/4$. Fig. 2 indicates that  the concurrence of two Rydberg impurities  exhibits periodical collapses and revivals during the time evolution. The initial concurrence can be completely revived during one evolution period. The peak value of the revivals is the initial amount of the concurrence.
It is interesting to note that the collapse and revival velocity of the concurrence  can be manipulated by changing initial-state parameter of the BEC $\alpha$. From Fig. 2 we can see that the collapse/revial velocity of the concurrence  speed up with the increase of  the initial-state parameter $\alpha$.

\subsection{The arbitrary pure state case}

We now study  quantum dynamics of the micro-micro entanglement when  the initial state of the BEC is a arbitrary pure states $|\varphi(0)\rangle$ while the two Rydberg impurities  initially is a Bell-type state $(\cos\theta|00\rangle+\sin\theta|11\rangle)$. Then, the initial states of the micro-macro system under our consideration can be written as
\begin{eqnarray}
|\psi(0)\rangle=(\cos\theta|00\rangle+\sin\theta|11\rangle)\otimes|\varphi(0)\rangle,
\end{eqnarray}

Making use of the Hamiltonian of the total system (7), we can find that at time $t\geq 0$  the state of the system becomes
\begin{eqnarray}
|\psi(t)\rangle&=&\cos\theta|00\rangle\otimes |\varphi'_0(t)\rangle +\sin\theta|11\rangle\otimes|\varphi'_1(t)\rangle,
\end{eqnarray}
where
\begin{eqnarray}
|\varphi'_0(t)\rangle &=&  e^{it\theta'_0(\hat{n})}|\varphi(0)\rangle, \\
|\varphi'_1(t)\rangle &=&  e^{it\theta'_1(\hat{n})}|\varphi(0)\rangle,
\end{eqnarray}
where we have introduced two running frequencies
\begin{eqnarray}
\theta'_0(\hat{n})&=&\omega-J -(\omega_b-\lambda)\hat{n}-\chi\hat{n}(\hat{n}-1), \\
\theta'_1(\hat{n})&=&-\omega-J -(\omega_b+\lambda)\hat{n}-\chi\hat{n}(\hat{n}-1).
\end{eqnarray}

From Eq. (24), it is easy to obtain the reduced density operator of two Rydberg impurities at time $t$
\begin{align}
\rho'_R(t)=&\cos^{2}\theta|00\rangle\langle00|+\sin^{2}\theta|11\rangle\langle11|\nonumber\\
&+\xi'(t)|00\rangle\langle11| +\xi'^{*}(t)|11\rangle\langle00|,
\end{align}
where $ \xi'(t)$ is defined by
\begin{equation}
 \xi'(t)=\frac{1}{2}\sin{2\theta}\langle\varphi'_1(t)|\varphi'_0(t)\rangle.
\end{equation}

The reduced density operator of two Rydberg impurities (29) can be expressed as the form of Eq.(19) with  $x(t)=0$ and $z(t)=\xi'(t)$. Hence from Eq. (20) we can obtain the quantum concurrence of the quantum state $\rho'_R(t)$  with the following expression,
\begin{eqnarray}
\mathcal{C}_1(t)&=&\mathcal{C}_1(0)\big|\langle\varphi'_1(t)|\varphi'_0(t)\rangle\big|, \nonumber \\
&=& \mathcal{C}_1(0)\big|\langle\varphi(0)|\exp\left(2it\lambda\hat{n}\right)|\varphi(0)\rangle\big|,
\end{eqnarray}
where we have used the following orthogonal relation
\begin{equation}
\langle\varphi'_1(t)|\varphi'_0(t)\rangle=\langle\varphi(0)|\exp\left[2it(\omega+\lambda\hat{n})\right]|\varphi(0)\rangle.
\end{equation}

From Eq. (31), we can see that the entanglement dynamics between two Rydberg impurities depends on the initial entanglement $\mathcal{C}_{1}(0)$, the initial state of the BEC $|\varphi(0)\rangle$, and the impurity-BEC interaction strength $\lambda$, even though the BEC is initially an arbitrary pure state.

\begin{figure}[htp]
\includegraphics[width=8.0cm,height=6.0cm]{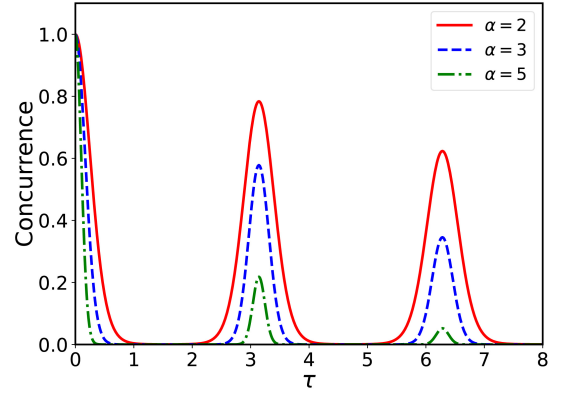}
\caption{Two Rydberg atoms concurrence versus the scaled time $\tau$ with different BEC initial coherent state parameter $\alpha=2,3,$ and 5. The BEC decoherence rates is assumed $\kappa=0.02$.}\label{fig3}
\end{figure}

\subsection{The decoherence influence}
No quantum system is totally isolated. Interactions between a quantum system and its environment cause quantum decoherence. Here we take into account the influence of the BEC decoherence on entanglement dynamics of two Rydberg impurities. The dynamic evolution of the density operator of totally system can be described by the zero-temperature master equation in the Born-Markov approximation
\begin{eqnarray}
\dot{\rho}=-\frac{i}{\hbar}[H,\rho]+\mathcal{L}(\rho),
\end{eqnarray}
where the superoperator $\mathcal{L}(\rho)$ is given by
\begin{eqnarray}
\mathcal{L}(\rho)=\kappa\Big(a\rho a^{\dag}-\frac{1}{2}a^{\dag}a\rho-\frac{1}{2}\rho a^{\dag}a\Big),
\end{eqnarray}
where $\kappa$ is the decay factor of the BEC, and $a$ is the BEC annihilation operator.

We now  numerically study influence of BEC decoherence on entanglement dynamics of two Rydberg impurities by the use of the  master equation (33). We focus on the investigation of effects of the BEC decoherence with the assistance of the Qu-TIP SoftWare Package \cite{36}. In Fig.3 we plot the dynamic evolution of the quantum concurrence of two Rydberg impurities with respect to the scaled time $\tau$ when the BEC decay factor $\kappa=0.02$, and two Rydberg impurities are initially in the Bell state $(|00\rangle + |11\rangle)/\sqrt{2}$ while the BEC is initially in a coherent state $|\alpha\rangle$ for different values of  the initial-state parameter of the BEC $\alpha=2, 3$, and $5$, respectively. From Fig. 3 we can see that the decoherence suppresses concurrence revivals and degrades the peaks of the concurrence in the dynamic evolution. The larger is the value of the initial-state parameter of the BEC, the faster the concurrence peaks decay.
Comparing Fig. 3 with Fig. 2, we can find that decoherence does not change the positions of the concurrence peaks, i.e., the time points at which the concurrence peaks appear in the processes of the time evolution for different the initial-state values of the BEC. However, the concurrence revivals will be asymptotically disappear in the long time evolution.

\section{\label{Sec:4} Complementarity relation between micro-macro and micro-micro entanglement }
In this section, we study the complementarity between micro-macro and micro-micro entanglement in the Rydberg impurities-doped  BEC system during the dynamic evolution. In to do so, we need to calculate micro-macro entanglement between two Rydberg impurities and the BEC. In what follows, we will demonstrate there exists an exact complementarity relation between micro-micro and micro-macro entanglement when two Rydberg impurities is initially in a Bell-type state while the BEC is initially in a coherent state and an arbitrary pure state, respectively.

\subsection{The coherent state case}

When the two Rydberg impurities  initially is a Bell-type state $(\cos\theta|00\rangle+\sin\theta|11\rangle)$ and the BEC is initially in a coherent state  $|\alpha\rangle$, at time $t$ the quantum state of the  Rydberg impurities-doped BEC is given by Eq. (11), which  is a micro-macro entangled state with two components. For a entangled pure state consisting of two components in the following form
\begin{eqnarray}
|\psi(t)\rangle=\frac{1}{N}[\mu|\eta\rangle\otimes|\gamma\rangle+\nu|\xi\rangle\otimes|\delta\rangle],
\end{eqnarray}
where $|\eta\rangle$ and $|\xi\rangle$ are normalized states of the first subsystem, $|\gamma\rangle$ and $|\delta\rangle$ are normalized states of the second subsystem  with complex $\mu$ and $\nu$. The normalization constant $N$ is given by
\begin{eqnarray}
N=\sqrt{|\mu|^2+|\nu|^2+ 2\textbf{Re}(\mu^*\nu p_{1}p_{2}^*)},
\end{eqnarray}
where we have introduced two overlapping functions
\begin{eqnarray}
p_1=\langle\eta|\xi\rangle, \hspace{0.5cm} p_2=\langle\delta|\gamma\rangle.
\end{eqnarray}

For the two-component entangled state given by Eq. (35), the quantum concurrence is given by  \cite{12}
\begin{eqnarray}
\mathcal{C}_2=\frac{2|\mu||\nu|}{N^2}\sqrt{(1-|p_{1}|^{2})(1-|p_{2}|^{2})}.
\end{eqnarray}

For the micro-macro entangled state (11), we have $N=1$, $\mu=\cos\theta$ and $\nu=\sin\theta$, two overlapping functions are given by
\begin{eqnarray}
p_1=0, \hspace{0.5cm}
p_2=\exp\Big\{2i\omega t+|\alpha|^{2}\big(e^{2i\lambda t}-1\big)\Big\}.
\end{eqnarray}

Making use of Eq.(38) and (39), we can obtain the concurrence of the micro-macro entangled state (11)  with the following expression
\begin{align}
\mathcal{C}_{2}(t)=&\mathcal{C}_{1}(0)\sqrt{1-\exp\Big\{2|\alpha|^{2}\big[{\rm cos}(2\lambda t)-1\big]}\Big\},
\end{align}
where $\mathcal{C}_{1}(0)$ is the initial micro-micro entanglement between two Rydberg impurities given by Eq. (21).

\begin{figure}[htp]
\includegraphics[width=8.0cm,height=6.0cm]{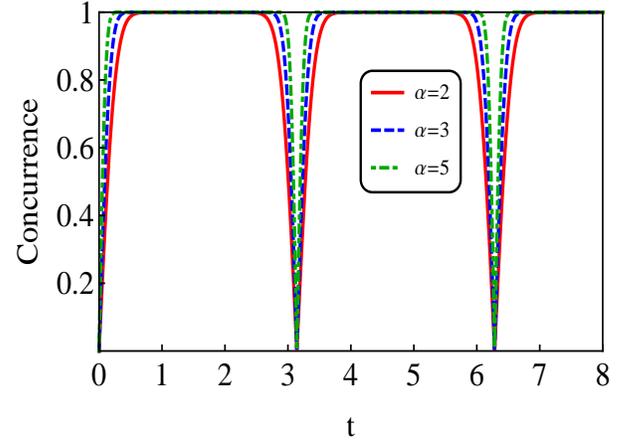}
\caption{Micro-Macro entanglement between two Rydberg impurities and the BEC denoted by the concurrence versus the scaled time $\tau$  for different values of the  initial-state parameter $\alpha$  with $\alpha=2,3$, and $5$, when the two Rydberg impurities is initially in the Bell state with $\theta=\pi/4$. }\label{fig4}
\end{figure}

Combining the concurrence of the micro-micro entanglement given by Eq.(21) and the expression of the micro-macro entanglement given by Eq.(40), we can find the following complementarity relation between the micro-micro  and the micro-macro entanglement
\begin{eqnarray}
\mathcal{C}_{1}^{2}(t)+\mathcal{C}_{2}^{2}(t)=\mathcal{C}_{1}^{2}(0).
\end{eqnarray}
Eq. (41) indicates entanglement transfer between impurities and impurities-BEC. The decreasing of $\mathcal{C}_{1}(t)$ will causes increasing of $\mathcal{C}_{2}(t)$. Inversely, The increasing of $\mathcal{C}_{1}(t)$ will causes decreasing of $\mathcal{C}_{2}(t)$.

The concurrence between Rydberg impurities and BEC is shown in Fig. 4 for different initial-state parameter of the BEC $\alpha$. Comparing Fig. 2 with Fig. 4, one can easily find the entanglement between Rydberg impurities and BEC will strengthen when the entanglement of two Rydberg impurities becoming weak even almost disappear. Then, entanglement of two Rydberg impurities will revives along with time evolution. At the same time, entanglement between Rydberg impurities and BEC will weaken until almost disappear. This phenomenon is a good description entanglement transfer between two Rydberg impurities and Rydberg impurities-BEC. The transfer of entanglement indicated that the quantum coherence or information transfer between two Rydberg impurities and Rydberg impurities-BEC.

In addition, no matter what we adjust $\lambda$ or $\alpha$, the concurrence of two Rydberg impurities will return to initial value after entanglement collapse for different coupling strength $\lambda$ or parameter $\alpha$. The reason of this phenomenon is that we don't consider decoherence of system.

\subsection{The arbitrary pure state case}

When the two Rydberg impurities  initially is a Bell-type state $(\cos\theta|00\rangle+\sin\theta|11\rangle)$ and the BEC is initially in an arbitrary pure  state  $|\varphi(0)\rangle$, at time $t$ the quantum state of the  Rydberg impurities-doped BEC is given by Eq. (24), which  is also a micro-macro entangled state with two components in the form of Eq. (35).

Making use of Eq.(35)-(37), for the micro-macro entangled state (24), we obtain $N=1$, $\mu=\cos\theta$ and $\nu=\sin\theta$, and two overlapping functions given by
\begin{eqnarray}
p_1=0, \hspace{0.5cm}
p_2=\langle\varphi'_0(t)|\varphi'_1(t)\rangle.
\end{eqnarray}

Hence, it is easy to get  the concurrence of the micro-macro entangled state (24)
\begin{eqnarray}
C_{Mi-Ma}=C_{Mi-Mi}(0)\sqrt{1-\big|\langle\varphi'_0(t)|\varphi'_1(t)\rangle\big|^{2}},
\end{eqnarray}
where $C_{Mi-Mi}(0)=|\sin(2\theta)|$ is the initial micro-micro entanglement of the Rydberg impurities-doped BEC system.

Combining the concurrence of the micro-micro entanglement given by Eq.(31) and the expression of the micro-macro entanglement given by Eq. (43), we can find the following complementarity relation between the micro-micro entanglement and the micro-macro entanglement
\begin{equation}
\mathcal{C}_{Mi-Ma}^{2}(t)+\mathcal{C}_{Mi-Mi}^{2}(t)=\mathcal{C}_{Mi-Mi}^{2}(0).
\end{equation}

Above complementarity relation between the micro-micro and micro-macro entanglement is the key result obtained in the present paper.
Eq.(44) indicates not only the  curious complementarity relation between the micro-micro and   micro-macro entanglement, but also reveals the physical mechanism for the generation of the micro-macro entanglement.
From Eq.(44) we can see that the initial  micro-micro entanglement $\mathcal{C}_{Mi-Mi}^{2}(0)$ is the seed entanglement to produce the  micro-macro entanglement under our consideration. Without the initial seed entanglement, the  micro-macro entanglement can not be generated in the system. On the other hand, From Eq. (44) we also see that the  micro-macro entanglement is produced through transferring from the micro-micro system of two Rydberg impurities  to the micro-macro system consisting of the Rydberg impurities and the BEC. The maximal amount of the micro-macro entanglement cannot be beyond the amount of the seed entanglement in the time evolution of the dynamics.

\subsection{The decoherence influence}

We now take into account the influence of the BEC decoherence on  the complementarity  between micro-micro and micro-macro entanglement in the micro-macro system. The dynamic evolution of the total density operator of the micro-macro system can be described by the zero-temperature master equation Eq. (33) in the Born-Markov approximation. In stead of the concurrence, we here use the negativity \cite{37,38} to  measure the  micro-micro and micro-macro entanglement  under BEC decoherence. The negativity for the  micro-macro system described by a density operator $\rho$ is given by the sum of the absolute values of the negative eigenvalues of the partially transposed density matrix $\rho^{pT}$,
\begin{equation}
N=\frac{1}{2}\sum_{j}(|\lambda_{j}|-\lambda_{j}),
\end{equation}
where the $\lambda_{j}$ is the eigenvalues of $\rho^{pT}$ and for a density operator of the micro-macro system $\rho=\sum_{klmn}p_{klmn}|k\rangle\langle l|\otimes|m\rangle\langle n|$ the partial transposed is given by $\rho^{pT}=\sum_{klmn}p_{klmn}|k\rangle\langle l|\otimes|n\rangle\langle m|$.

\begin{figure}[htp]
\includegraphics[width=8.0cm,height=6.0cm]{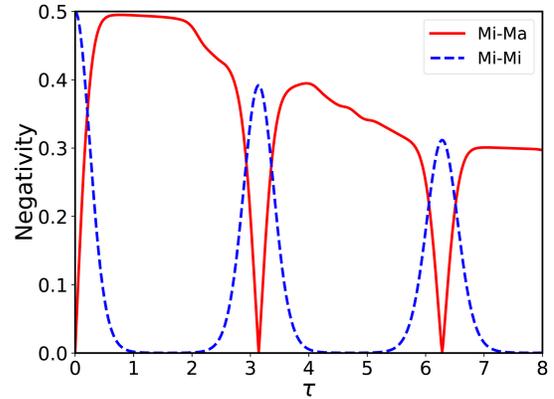}
\caption{Micro-Macro entanglement (red solid line) between two Rydberg impurities and the BEC denoted by the negativity versus the scaled time $\tau$, Micro-Micro entanglement (blue dashed line) between two Rydberg impurities denoted by the negativity versus the scaled time $\tau$, when the two Rydberg impurities is initially in the Bell state with $\theta=\pi/4$ and initial state parameter of the BEC $\alpha=2$. The BEC decoherence rates is assumed $\kappa=0.02$.}\label{fig5}
\end{figure}

In Fig.5 we have plotted the dynamic evolution of the negativity to describe micro-macro  and micro-micro entanglement with respect to the scaled time $\tau$ when  two Rydberg impurities are initially in the Bell state $(|00\rangle + |11\rangle)/\sqrt{2}$ while the BEC is initially in a coherent state $|\alpha\rangle$ with $\alpha = 2$ for  the BEC decay factor $\kappa = 0.02$.  In Fig. 5 the solid and dashed lines denote the  negativity for micro-macro  and micro-micro entanglement, respectively.  From Fig.5 we can see that the complementarity relation between the micro-micro  and   micro-macro entanglement in the micro-macro system is generally preserved in the time evolution process even though in the presence of the BEC decay. In particular, we can find that the peaks of the micro-micro negativity exactly correspond to the dips of the micro-macro negativity although  the BEC decoherence suppresses the entanglement creation.

\section{\label{Sec:6} conclusion }

In this paper, we have studied the complementarity between micro-micro and micro-macro entanglement based on a  micro-macro system which consists of two microscopic Rydberg impurities and the macroscopic BEC. We investigate dynamics of quantum entanglement in  a Bose-Einstein condensate (BEC) system with two Rydberg impurities.  It is found that strong micro-macro  entanglement  between Rydberg impurities and the BEC can be generated by the use of  initial micro-micro entanglement between two Rydberg impurities, which acts as the seed entanglement to create micro-macro entanglement. It is shown that the micro-macro system exhibits not only micro-micro entanglement collapse and revival (ECR) between two Rydberg impurities, but also micro-macro ECR between  Rydberg impurities and the BEC. We point out the possibility of controlling the ECR through changing key parameters of the impurities-doped BEC system such as initial-state parameters and Rydberg impurities-BEC coupling strength, and uncover the physical mechanism behind the ECR phenomenon. We demonstrate a curious complementarity relation between the micro-micro entanglement and  the micro-macro entanglement, and find  that micro-micro entanglement can perfectly transfer into the micro-macro entanglement. We have numerically study the effect of the BEC decoherence on the complementarity.  It is indicated that the complementarity property can be sustained to some extend even though in the presence of the BEC decay. Our results not only  cast a new light on the complementarity in quantum mechanics, but also provide a route for understanding and controlling quantum entanglement in micro-macro quantum systems.

\acknowledgments This work was supported by the National Natural Science
Foundation of China under Grants No. 11775075 and No. 11434011.

\end{document}